\documentclass[prl, aps,twocolumn, amsmath, graphicx, latexsym, asmsymb, amsfonts]{revtex4}

\usepackage{epsfig}

\begin{document}
\title{Regional versus Global entanglement in Resonating-Valence-Bond states}
\author{Anushya Chandran\(^{1,2}\), Dagomir Kaszlikowski\(^2\), Aditi Sen(De)\(^3\), Ujjwal Sen\(^3\), and Vlatko Vedral\(^{2,4}\)}
\affiliation{\(^1\)Department of Electrical Engineering, The Indian Institute of Technology Madras, Chennai, India\\
\(^2\) Department of Physics, National University of Singapore, 117542 Singapore, Singapore\\
\(^3\)ICFO-Institut de Ci\`encies Fot\`oniques, Mediterranean Technology Park,
08860 Castelldefels (Barcelona), Spain\\
\(^4\)The School of Physics and Astronomy, University
of Leeds, Leeds, LS2 9JT, United Kingdom}


\begin{abstract}
We investigate the entanglement properties of resonating-valence-bond states on two and higher dimensional lattices, 
which play a significant role in our understanding of various many-body systems. We show that these states are genuinely multipartite entangled, 
while there is only a negligible amount of two-site entanglement. We comment on possible physical implications of our findings.
\end{abstract}

\maketitle

\emph{Introduction.}
In quantum many-body physics, resonating-valence-bond (RVB) states have received a lot of attention due to its  importance in 
the description of different 
phenomena. 
They are used to describe the 
resonance of covalent bonds in organic molecules, behavior of Mott insulators without long-range  antiferromagnetic order \cite{anderson}, 
d- and s-wave superconducting states \cite{zoller}, superconductivity in organic solids \cite{organic-solids}, and the recently discovered 
insulator-superconductor transition in boron-doped diamond \cite{boron}. There are many other applications of RVB states (see e.g. \cite{review}). 
Moreover, RVB states have been suggested as a basis for fault-tolerant 
topological quantum computation \cite{kitaev}. We believe that successful applications of RVB states partially rest on the interesting 
entanglement properties that they have, and this particular aspect has not received much attention in the literature. 

Various tools of quantum information (QI) have been successfully employed to understand many-body systems \cite{pisa-review}. 
In particular, entanglement has been found to be an indicator of quantum phase transitions \cite{fazio-nielsen}. Moreover, 
condensed matter systems can be efficiently simulated using techniques related to entanglement \cite{vidal}. The usefulness 
of entanglement in condensed matter physics leads us to consider it in the context of the RVB states.  

\emph{The main thesis and results.}
The main thesis of this paper is that the RVB states have a very particular structure from the point of view of the distribution 
of entanglement. More specifically, entanglement stretches over the significant fraction of the lattice, while there is virtually 
no entanglement when we restrict ourselves to small regions of the lattice. This fact may play a significant role in the physics 
of the RVB states.

We show that the most general RVB-type states on the two- (or more) dimensional
 lattice do not contain a significant amount 
of bipartite entanglement (BE) between any two sites of the lattice. However, genuine multipartite entanglement is present 
when we consider the whole lattice. We exemplify our results by considering two extreme cases: the so-called RVB gas and RVB liquid. 

Among the QI concepts that we use to prove these results, are ``monogamy of entanglement'' \cite{monogamy2}, which
places restrictions on BE in a multipartite scenario,
and ``quantum telecloning'' \cite{telecloning}, a phenomenon that 
uses multiparty entanglement to produce approximate copies (clones) of a given state at separated locations. 
Surprisingly, it turns out that one can obtain more precise estimations of entanglement by using quantum telecloning, rather than by monogamy.  

\emph{Derivations and discussions.}
Let us begin with a brief formal definition of entangled and separable states. A pure state of two parties is said to be entangled (separable) 
if it cannot (can) be expressed
as a tensor product of two pure states at the two parties. An entangled (separable) mixed state of two parties is one which cannot (can)
be expressed as a probabilistic mixture of separable pure states. 
Lastly, a pure state of an arbitrary number of parties is said to be  
genuinely multiparty entangled, if it is not separable in any bipartite splitting. We will not have occasion to consider 
further general scenarios.

For definiteness, we will state and derive our results for any two-dimensional (2D) lattice (including infinite ones).
However, it will be apparent that most of our considerations can be carried over to higher dimensions.
Each lattice site is occupied by a qubit (a two-dimensional quantum system, e.g. a spin-1/2 particle).

Consider a 2D lattice that is a union of two sub-lattices, $A$ and $B$, where any site from 
sub-lattice $A$ (\(B\)) does not have any sites from the same sub-lattice as its nearest neighbors (NNs).
An RVB state
on such a  lattice is 
\cite{Anderson-RVB}
\begin{equation}
\label{abir}
|\psi\rangle =\sum
h(i_1,\dots,i_N,j_1,\dots,j_N)
|(i_1,j_1)\dots(i_N,j_N)\rangle, \nonumber
\end{equation}
where 
the sum runs over \(i_{\alpha}\in A\), and \(j_{\beta}\in B\), 
$N$ is the number of sites in each  
sub-lattice,
and  $|(i_k,j_k)\rangle$ denotes the singlet (\emph{dimer}), 
$\frac{1}{\sqrt{2}}(|\frac{1}{2}\rangle_{i_k}|-\frac{1}{2}\rangle_{j_k}-|-\frac{1}{2}\rangle_{i_k}|\frac{1}{2}\rangle_{j_k})$,
connecting a site in  $A$ with a site in  $B$. 
The function $h$ is only assumed to be isotropic over the lattice.
(The original definition, e.g. in Ref. \cite{Anderson-RVB}, is far more restrictive, in that \(h\) is assumed 
to be positive, factorisable, and only a function of 
the distance between the sites.)
Every element in the superposition in \(|\psi\rangle\) 
 is said to be a covering of the lattice under consideration. Such an RVB 
state can be defined on any lattice and in any dimension. 
It is useful  to consider two extreme examples:
RVB 
gas and  RVB liquid. 
The gas is the RVB state where the function $h$ is a constant (so that the state is made up from coverings of equal strength), 
whereas for the liquid, we consider those coverings that contain 
only NN dimers. 

Let us first start with an observation on the rotational properties of the reduced density matrices of $|\psi\rangle$, which is important in investigations of entanglement properties of the state $|\psi\rangle$. We notice 
that any $n$-body density matrix $\rho_{k_1,\dots,k_n}$ describing $n$ arbitrary sites $k_1,\dots,k_n$ (irrespective
of their distribution among the sub-lattices, and including the case when all are from a single sub-lattice)
is rotationally invariant (i.e. invariant under the action of \(U^{\otimes n}\), where \(U\) is a general unitary acting on the 
qubit Hilbert space). 
This is a consequence of the rotational invariance of $|\psi\rangle$, which is  a superposition of the 
rotationally invariant singlets.
The proof of this fact in the case of a two-body density matrix (easily extendible to  an arbitrary number of  sites) is as follows:
\begin{eqnarray}
&&\rho_{ij}=\sum_{\overline{ij}}|\langle \overline{ij}|\psi\rangle|^2=\sum_{\overline{ij}}|\langle \overline{ij}|
U^{\otimes 2N}|\psi\rangle|^2\nonumber\\
&&=U^{(i)}U^{(j)}\sum_{\overline{ij}}|\langle\overline{ij}|\psi\rangle|^2(U^{(i)}U^{(j)})^{\dagger}.
\end{eqnarray}
Here \(\Lambda^{(k)}\) denotes the operator \(\Lambda\) at the site \(k\), the summation excludes the $i$th and $j$th site, and 
$\langle \overline{ij}|\psi\rangle$ is the partial scalar product.
The rotational invariance implies, in particular,  that any single-site density matrix is in a completely 
depolarized state, and any two-body density matrix is a ``Werner state'' 
\cite{Werner} 
\(\rho_W(p)=p|(i,j)\rangle\langle (i,j)|+(1-p)I_4/4\),
with $-\frac{1}{3}\leq p\leq 1$ and $I_4$ is an identity operator for two spins.

Let us now investigate the entanglement properties of the state $|\psi\rangle$.
We begin by analyzing 
the entanglement properties of any two-site density matrix. The first tool we are going 
to use is the so-called monogamy of entanglement \cite{monogamy2}. In short,  monogamy places 
restrictions  on the amount
 of entanglement that a certain quantum system can have with another, given that the former is 
already entangled with a third system.
For instance, if two systems are maximally 
entangled, this entirely excludes entanglement between any of them and some  other system. However, if the two systems are not maximally 
entangled, this does not exclude entanglement with the third one.

It is possible to quantify the notion of monogamy in terms of the ``tangle'' \cite{tangle}.
We will only have occasion to 
consider states of a qubit and a \(d\)-dimensional quantum system (qudit). 
The tangle for a pure state \(|\phi\rangle_{\mathbb{A}\mathbb{B}}\) of a qubit (\(\mathbb{A}\)) and 
a qudit (\(\mathbb{B}\)) is a measure of quantum correlation (entanglement), and 
is defined as \(\tau(|\phi\rangle) = S_L(\mbox{tr}_\mathbb{B}(|\psi\rangle\langle\psi| ))\), where the linearized entropy 
\(S_L (\varrho) = 2(1-\mbox{tr}(\varrho^2))\). For a mixed state \(\eta_{\mathbb{A}\mathbb{B}}\), the tangle 
is defined by the convex roof construction: \(\tau(\eta) = \inf_{p_x, |\phi_x\rangle} \sum_x p_x \tau(|\phi_x\rangle)\), where 
the infimum is over all probabilistic pure-state decompositions, \(\sum_x p_x |\phi_x\rangle \langle \phi_x|\), of \(\eta\).  
For a state \(\eta\) of two qubits, the tangle is given by the square of \(\max(0,\lambda_1-\lambda_2-\lambda_3-\lambda_4)\), where \(\lambda_i\) 
are the square roots of the eigenvalues, in decreasing order, of \(\eta \tilde{\eta}\), 
with \(\tilde{\eta}= \sigma_y \otimes \sigma_y \eta^* \sigma_y \otimes \sigma_y\),
the complex conjugation being performed in the \(\sigma_z \otimes \sigma_z\) basis.
In this paper, \(\vec{\sigma}= (\sigma_x, \sigma_y, \sigma_z)\), where \(\sigma_\alpha\) are the Pauli matrices. 

The monogamy of entanglement for a state \(\rho_n\) of \(n\) qubits \(1,2, \dots, n\) can be 
quantified by the inequality
\(\sum_{k=2}^n\tau(\rho_{1k})\leq \tau(\rho_{1:(2\dots n)})\),
where $\tau(\rho_{1k})$ denotes the tangle between qubits $1$ and  $k$, and $\tau(\rho_{1:(2\dots n)})$, the tangle 
between qubit $1$ and the aggregate of all the other qubits $2,3,\dots, n$ treated as a single
($2^{n-1}$-dimensional) quantum system \cite{monogamy2}. In general, \(\tau\) can vary between 0 and 1, but  monogamy
constrains the entanglement (\(\tau\)) that the particle 1 can have with each of 
\(2,3,\dots,n\).

We now use the monogamy constraint to 
estimate two-site entanglement in an RVB state. 
For definiteness, let us consider a 2D square lattice, and let us choose an arbitrary site \({\cal A}\) on the lattice. 
To focus attention, we assume that \({\cal A}\) belongs to the sub-lattice $A$. The site \({\cal A}\) 
has four NNs, say $B_1,B_2,B_3$ and $B_4$, belonging to the sub-lattice $B$. As noted before, 
each pair $({\cal A},B_k)$, 
is in a Werner state, with the \emph{same} $p$, the last fact being due to the assumption of the isotropic nature
of the RVB state over the lattice. 
If the pair $({\cal A},B_k)$ is entangled, i.e. $p> \frac{1}{3}$ \cite{pereshoro}, its tangle reads 
\(\tau(\rho_{{\cal A}B_k})=(3p-1)^2 /4\).
The tangle $\tau(\rho_{{\cal A}:(B_1B_2B_3B_4)})$, between the site ${\cal A}$, 
and its NNs (treated as a single  $2^{4}$-dimensional system)
 cannot be 
greater than one.
Therefore, monogamy of entanglement gives us our first upper bound on $p$, for any pair of NNs:
\(p\leq \frac{2}{3}\).
Of course, this upper bound does not tell us if there really is any
entanglement between the NNs. However, we know that this is 
a weak bound because of the imprecise estimation of the tangle $\tau(\rho_{{\cal A}:(B_1B_2B_3B_4)})$.
As we show later, this bound can be improved by 
using some additional techniques from QI theory. 

The above reasoning 
can be applied to pairs of sites that are far away from each other, resulting, in general, in stronger bounds. E.g.,
if there are $R$ sites at the distance $r$ from the site ${\cal A}$, the monogamy inequality gives us 
\(p\leq \frac{1}{3}+\frac{2}{3\sqrt{R}}\),
where now $p$ refers to the Werner state between the site ${\cal A}$, and any site at the distance $r$ from ${\cal A}$. 
The number of equidistant points increases proportionally to
$r$, suppressing any possible entanglement between such sites. 
Similar techniques can be used for other lattice geometries and other dimensions.

We  now demonstrate that a different approach, based on the phenomenon of (approximate) quantum telecloning \cite{telecloning},
gives more stringent bounds on the  
amount of entanglement shared between pairs of sites.
Briefly, the telecloning phenomenon composes two concepts of QI: ``quantum teleportation'' \cite{teleportation}, which 
transfers  a quantum state from one location to another by using shared entanglement and a small 
amount of classical communication, and ``quantum cloning'' \cite{cloningRMP}, which deals with the production of approximate copies 
of a given unknown quantum state. In telecloning, the approximate copies of the given unknown state are produced at separated locations, by using a 
shared multipartite entangled state, along with classical communication. 

To use the telecloning results for our purpose, we again consider a site ${\cal A}$ 
surrounded by four equidistant NNs $B_1,B_2,B_3,B_4$. By attaching an auxiliary qubit to the qubit 
at site ${\cal A}$, performing the Bell measurement (measurement projecting onto 
the singlet and the triplets) on this joint two-qubit system,
 and broadcasting the resulting two bits of classical information, we can quantum teleport \cite{teleportation} an arbitrary state 
of the auxiliary qubit to the neighbors $B_k$, with a certain (non-unit) fidelity, where 
the fidelity of a process with input \(|\phi\rangle\) and output \(\varrho_{\phi}\) is defined as 
\(
\int \langle \phi | \varrho_{\phi} |\phi\rangle d\phi,
\)  
with \(d\phi\) being the unitarily invariant measure on the input space.
This is exactly what is achieved in quantum telecloning, although the shared state that was used for the purpose was different from ours
\cite{telecloning}.
Due to isotropy of 
$|\psi\rangle$,  the fidelity of teleportation, 
$F_{tele}$, to the four sites is the same, and is
\(F_{tele} = \frac{p+1}{2}\) \cite{Horodecki}.
This fidelity cannot exceed the fidelity of the optimal symmetric cloning, $F_{clone}$, producing four copies of the initial state. 
The 
optimal quantum cloning machine that produces 
$M$ copies from a single copy of the input qubit leads to the fidelity 
\(F_{clone} = \frac{2 M+1}{3 M}\) \cite{cloningRMP}.
%
Therefore an upper bound of $p$, for NNs, can be obtained from the inequality $F_{tele}\leq F_{clone}$
 and it reads 
\(p\leq \frac{1}{2}\).
This bound is much better than the one obtained from the monogamy argument. 
Now $p=\frac{1}{2}$ implies 
a very low  
entanglement of formation ($\approx 0.023$ ebits) \cite{tangle}.
The entanglement of formation \cite{qwertyuiop} of a 
two-party pure state is the asymptotic ratio of the number of singlets (ebits) that is required to prepare the state by local 
quantum operations and classical communication. The generalization to mixed states is again done via the convex roof construction, discussed
before.
We therefore have a strong indication that there is virtually no bipartite
entanglement between any two sites on the lattice.

As in the case of the monogamy, one gets tighter bounds for the entanglement 
between ${\cal A}$ and the equidistant qubits at distance $r$, because, in general, more clones are 
 formed with increasing $r$. 
Using 
$F_{tele}\leq F_{clone}$, we obtain
\(p\leq \frac{1}{3}+\frac{2}{3R}\). 
This is a \emph{square-root improvement} (over the bound obtained from monogamy) 
in the convergence to the separability point \(p=1/3\).

Note 
that telecloning is viewed here as a monogamy of the amount of QI that 
can be sent (teleported) in a distributed network, while the original monogamy of entanglement \cite{monogamy2} was a constraint 
on the shared entanglement in a network. To understand this, we remember that shared entanglement is a resource for 
sending QI \cite{teleportation}. What is curious is that telecloning seems to point towards a more stringent monogamy, than the 
ones already known, even though its original purpose was not at all related to sharing of entanglement.

Let us now consider the two special cases mentioned before: the RVB gas and the RVB liquid. 
For the 
RVB gas, 
since any pair of sites from different sublattices has the same \(p\), 
monogamy of entanglement gives us the strong bound 
\(p\leq \frac{1}{3}+\frac{2\sqrt{2}}{3\sqrt{N}}\),
where $N$ is the number of sites in each sub-lattice.
The bound obtained from the 
telecloning argument is tighter: 
\(p\leq \frac{1}{3} +\frac{2}{3 N}\). 
Both bounds predict separability in the case of an infinite lattice.
Interestingly, direct computation for RVB gases of size 6 and 8 saturates the telecloning bounds.

The situation is much more complicated in the case of the RVB liquid. 
By the techniques used here, one cannot obtain any tighter bounds than the ones already presented. 
However, one can get some additional information on the structure of the BE between the sites using the standard techniques 
from condensed matter physics, which we now briefly describe.
In condensed matter physics, one is usually interested in the behavior of the correlation function (CF) 
between two sites $i$ and $j$, $\langle\psi|\vec{S}_i\cdot\vec{S}_j|\psi\rangle$, 
where \(\vec{S}= \frac{1}{2}\vec{\sigma}\). 
Ref. \cite{tasaki} shows
 that the 
two-point CF can be computed by using the so-called loop coverings. A brief explanation of the method is as follows. The 
state $|\psi\rangle$ in the case of the 
RVB liquid can be written simply as $|\psi\rangle=\sum_k |c_k\rangle$, 
where $|c_k\rangle$ represents a certain configurations of dimers 
between NNs. To compute the two-point CF, 
one needs to know $\langle c_k|\vec{S}_i\cdot\vec{S}_j|c_l\rangle$, for an arbitrary $k$ and $l$. Each pair of the kets, 
$\{|c_k\rangle,|c_l\rangle\}$,  can
be graphically represented as lines (bonds) between pairs of sites on the lattice. These bonds can form two kind of 
non-overlapping loops:
degenerate and non-degenerate. Degenerate loops encircle two neighboring sites, and non-degenerate ones join more than two sites such that each site belongs to only one loop.
The evaluation of the expression $\langle c_k|\vec{S}_i\cdot\vec{S}_j|c_l\rangle$ is very simple: it is zero if $i$ and $j$ belong to 
two different loops, and it is proportional 
to $\pm\frac{3}{4}$ if $i$ and $j$ belong to the same loop. We must take
the plus sign if $i$ and $j$ belong to different sub-lattices, and 
minus sign otherwise.
Using the above concepts, 
one arrives at the formula
$
\langle\psi|\vec{S}_i\cdot\vec{S}_j|\psi\rangle = (-1)^{|i-j|} \frac{3}{4}\frac{\sum_g X(i, j)4^{n(g)}2^{d(g)}}{\sum_g4^{n(g)}2^{d(g)}},
$
 where the summation is over all 
graphs created by the dimer coverings, and $(-1)^{|i-j|}$ equals to $+1$ if $i$ and $j$ belong to 
different sub-lattices, and to $-1$ otherwise.
 The function $X(i,j)$ is $1$ if $i$ and $j$ belong to a loop, and is zero otherwise.
 The importance of the above equation, 
for this paper, stems from the fact that 
$\langle\psi|\vec{S}_i\cdot\vec{S}_j|\psi\rangle$ is exactly equal to the parameter \(p\)
in the Werner state describing the reduced 
 density matrix of the sites $i$ and $j$. Therefore, 
for sites from the same sub-lattice, $p$ is either strictly 
negative or zero (zero only if the denominator grows faster than the numerator), which excludes entanglement between such sites.

By using the above method, we have found that for the RVB liquid, any two NN sites in the interior of a square \(4 \times 4\) lattice,
\(p \approx 0.2004\), which interestingly corresponds to a separable state. Based on this fact, it is reasonable to assume that 
this separability is not affected by increasing the lattice size, confirming our thesis of having no two-site entanglement in an RVB.
Higher level entanglement of course exists, as we will show 
below.



Note 
here 
that 
the concept of quantum
telecloning gives  upper bounds on the long-range behavior of the two-point 
CFs for an arbitrary RVB state on a lattice (even three-dimensional ones). To our knowledge, 
this is the first instance when such a connection is observed.  
This is an example where techniques from QI can be applied to deal with 
phenomena that are interesting in 
condensed matter physics.

Let us now consider the multipartite entanglement properties of an arbitrary RVB state $|\psi\rangle$.
We begin by observing that any odd number of sites, of an arbitrary RVB state, is entangled to the rest of the lattice. To prove this,
 it is enough to show that any such arbitrary odd number of qubits
 is in a mixed
state. (Note that the whole state is pure.) 
This however follows from the rotational invariance of the density matrix that describes the odd number of qubits, 
as there is no pure state of 
an odd number of qubits that is rotationally invariant.  
Therefore, any set of an odd number of qubits is entangled to 
the rest of the lattice. In particular, any single qubit is maximally entangled with the rest of the lattice.

To show that a certain RVB state has genuine \(2N\)-party multiparty entanglement, we are left with 
showing that any set of an even number of sites is entangled to the rest of the lattice. 
First, consider the RVB state in a bipartite splitting between any two sites of the lattice and the remaining part of it.
As we have seen before, such a state is in a Werner state, with \(p\leq 1/2\). In particular, the state is not pure. Therefore,
any two sites of the lattice is entangled to the rest of the lattice.

Consider now any even subset of the lattice consisting of the sites $e_1,\dots,e_n$ ($n$ is even).
Suppose that this subset is not entangled to the rest of the lattice, i.e.
the state $|\psi\rangle$ can be written as $|\psi\rangle = |\psi_{e_1\dots e_n}\rangle |\phi_e\rangle$. 
As the
function \(h\)
defining $|\psi\rangle$
is isotropic, there  exists a subset $f_1,\dots f_n$ having one common site with the subset $e_1,\dots,e_n$, 
say $e_1=f_1$, such that $|\psi\rangle=|\psi_{e_1f_2\dots f_n}\rangle |\phi_f\rangle$. However,
this means that the qubit at the site $e_1$ must be disentangled from the rest of the lattice, 
which is not possible because every qubit on the lattice is maximally entangled to the rest of the lattice as shown before.
In this proof, we have assumed that either the lattice is infinite, or that it has periodic boundary conditions.
It is worthwhile to note that numerical simulations in Ref. \cite{Anderson-RVB}  indicate that any two-site state is a Werner state with nonzero 
\(p\), in the case of RVB states with 
factorisable, nonnegative \(h\), depending only on the distance between the lattice sites connected by the dimers in the corresponding covering. 

\emph{Conclusions.} 
We have shown that isotropic resonating valence bond states in any two or higher dimensional 
lattice have only an insignificant amount of two-site entanglement, while having genuine multi-party entanglement. 
To understand this, it is tempting to point to the large number of inter-site connections in the terms that build up the 
RVB state, which, intuitively, would result in genuine 
multiparty entanglement, while precluding any two-site entanglement due to the monogamous nature of entanglement.
However, one should be cautious, as counterexamples exist (e.g. \cite{Brijesh_Kumar_Zindabad}). Traditionally, 
properties of many-body systems are mostly quantified using bipartite measures such as two-point correlation functions, 
concurrence, block entropy, to name a few. The present work shows that the RVB structure is far richer, and may therefore require 
more elaborate ways of quantifying its properties. On the quantum computational side, this intricate structure may allow for different 
ways of information processing, such as coherent broadcasting of qubits \cite{telecloning}.  
Finally, 
our results also apply to the ground state of a three dimensional antiferromagnetic 
Heisenberg model with nearest neighbor interactions and a possible next-nearest neighbor ferromagnetic term \cite{Marshall}.

AS and US thank the National University of Singapore for their hospitality.
We acknowledge support from 
DFG 
(SFB 407, SPP 1078, SPP 1116), 
ESF QUDEDIS, Spanish MEC (FIS-2005-04627, Consolider Project QOIT, \& Ram{\'o}n y Cajal), EU IP SCALA, QIT strategic grant R-144-000-190-646,
Engineering and Physical Sciences Research Council, UK, and Royal Society, UK.


\end{document}